\documentclass[aps,prl,twocolumn,showpacs,superscriptaddress,groupedaddress]{revtex4}  
\usepackage{graphicx}        
\usepackage{dcolumn}         
\usepackage{bm}              
\usepackage{amssymb}         
\usepackage{amsmath}         
\usepackage{gensymb}         
\usepackage{upgreek}         
\usepackage{numprint}        
\usepackage{array,makecell}  
                            
\usepackage{color} 
\usepackage{ulem}  
\begin{document}

\widetext

\title{Probing the viscoelastic properties of polyacrylamide polymer gels in a wide frequency range.}
\author{Y.~Abidine} \affiliation{Univ. Grenoble Alpes, LIPHY, F-38000 Grenoble, France} \affiliation{CNRS, LIPHY, F-38000 Grenoble, France} 
\author{V.~M.~Laurent} \affiliation{Univ. Grenoble Alpes, LIPHY, F-38000 Grenoble, France} \affiliation{CNRS, LIPHY, F-38000 Grenoble, France}
\author{R.~Michel} \affiliation{Univ. Grenoble Alpes, LIPHY, F-38000 Grenoble, France} \affiliation{CNRS, LIPHY, F-38000 Grenoble, France}
\author{A.~Duperray} \affiliation{INSERM, IAB, F-38000 Grenoble, France} \affiliation{Univ. Grenoble Alpes, IAB, F-38000 Grenoble, France } 
\affiliation{CHU de Grenoble, IAB, F-38000 Grenoble, France }
\author{L. I.~Palade} \affiliation{Universit{\'e} Lyon, CNRS, Institut Camille Jordan, UMR 5208, INSA-Lyon, P{\^o}le de Math{\'e}matiques, F-69621 Villeurbanne, France}
\author{C.~Verdier\footnote{claude.verdier@ujf-grenoble.fr}} \affiliation{Univ. Grenoble Alpes, LIPHY, F-38000 Grenoble, France} \affiliation{CNRS, LIPHY, F-38000 Grenoble, France}
\vskip 0.25cm

\date{\today}

\begin{abstract}

Polymer gels have been shown to behave as viscoelastic materials but only a small amount of data is usually provided in the glassy transition. 
In this paper, the dynamic moduli $G'$ and $G''$ of polyacrylamide hydrogels are investigated using both an AFM in contact force modulation mode and a 
classical rheometer. The validity is shown by perfect matching of the two techniques. 

Measurements are carried out on gels of increasing polymer concentration in a wide frequency range. A model based on fractional derivatives is proposed, 
covering the whole frequency range. $G_N^{0}$, the plateau modulus, as well as $n_f$, the slope of the $G''$ modulus, are obtained at low frequencies. 
The model also predicts the slope $a$ of both moduli in the transition regime, as well as a new transition time $\lambda_T$. The whole frequency spectrum 
is recovered, and the model parameters contain interesting information about the physics of such gels.
\end{abstract}


\maketitle

Polymers exhibit interesting rheological behavior in the sense that they can successively behave as liquids, elastic materials showing a rubbery 
plateau, then undergo a glassy transition before reaching the solid domain \cite{larson1999}. These processes are temperature-dependent.
Because of this broad range of properties, polymers are widely used in industrial applications, as well as biological processes. However, it is often 
difficult to characterize their material properties, as the range of frequencies involved covers several decades \cite{Baumgaertel1992,palade1996}. 
Techniques such as classical rheometry, diffusing-wave spectroscopy, dynamic light scattering \cite{dasgupta-Weitz2005} or ultrasonic experiments 
have been used to describe the complex behavior of polymers each in its own range of frequencies \cite{verdier1998a,Longin1998}. In particular, an important way 
to extend the linear viscoelastic behavior (LVE) is found by using the time--temperature superposition principle, wherein results obtained at various 
temperatures are shifted onto a reference temperature master curve \cite{palade1996}. These observations have motivated quite a lot of theoretical studies. 
Different models providing relaxation spectra have been proposed, ranging from multiple Maxwell models to continuous relaxation 
spectra \cite{Baumgaertel1992}, involving both liquid and glassy modes. The concept of soft glassy rheology \cite{sollich1997,sollich1998} appeared 
recently and provides another alternative suited for many systems. Indeed it is based on the idea that sub--elements in the microstructure are 
linked via weak interactions, and are in a disordered metastable state. Based on this concept, many complex fluids can be described thanks to this 
model, in particular packed colloidal suspensions, the cell cytoskeleton \cite{Fabry2001} as well as foams or slurries.

Due to their cross-linked network, polymer gels share similar properties \cite{sollich1997} with polymers. They can be characterized using modern microrheology 
techniques \cite{Mason1997,Crocker2000}, as applied in particular for actin networks \cite{Chaudhuri2007,Dalhaimer-Discher2007}. 
The behavior of classical gels is in fact similar in the glassy transition domain, but no modelling attempt has been made so far to characterize the 
entire frequency domain covered by recent instruments. Therefore, it is interesting to characterize a wide domain of frequency for various 
polymeric gels, especially biological systems, and develop a model for such behavior. This is the main purpose of the work presented here. In addition, 
a new AFM--based microrheology method \cite{Alcaraz2003,Hiratsuka2009,Abidine2013} will be used allowing to investigate a wide range of frequencies, in combination
with classical rheometry. This technique to probe the mechanical properties of biological cells locally using dynamic AFM measurements was developped \cite{Alcaraz2003}, 
but has not yet been validated on a model system, like a polyacrylamide gel. Here, we choose to characterize the behavior of polyacrylamide hydrogels 
due to their interesting mechanical properties depending on the cross-linked network. These gels are known to exhibit a viscoelastic behavior, 
with an elastic modulus $G'$, and a frequency--dependent viscous modulus $G''$, usually one decade below in the classical rheology 
domain $[0.01\,\mathrm{Hz}-10\,\mathrm{Hz}]$ \cite{ambrosi2009}. The elastic modulus ($G_N^{0}$) has been investigated and increases with 
acrylamide concentration \cite{Hecht1978}. Thus, changing the cross-linked network and measuring the dynamical moduli in a wide range of 
frequencies can bring forward new data related to the physical properties of the gels. 

Polyacrylamide gels were synthesized by mixing acrylamide ($30\%$ w/w) at four different weight concentrations ($5-7.5-10-15\%$), 
and N,N’--methylene--bisacrylamide $1\%$ w/w at a fixed concentration $0.03\%$ in deionized water. This means that the hydrogels were slightly crosslinked. 
Polymerization was then initiated by incorporating N,N,N’,N’-tetramethylethylenediamine (TEMED, Sigma) and ammonium persulfate $10\%$ solution (APS), 
as described in \cite{Pelham1999}. Gels of thickness $70\,\upmu \mathrm{m}$ were prepared on a pre--treated glass Petri dish for a better  
adhesion \cite{Pelham1999}. Gels were always kept in humid conditions, so that they were swollen and in equilibrium. They 
were set onto an AFM (JPK Instruments, Berlin) equipped with an inverted microscope (Zeiss, model D1, Berlin).
The AFM chips (Bruker, MLCT, pyramid shape, tip half--angle $\theta=20\degree$) were mounted onto the AFM glass block and calibrated using the thermal 
fluctuations method. Then an initial indentation $\delta_0$ of the sample was made under a prescribed force $F_0$ given by Hertz model:
\begin{equation}
F_0 = \frac{3\,E \tan\theta}{4\,(1-\nu^2)}\delta_0^2
\label{indentation}
\end{equation}
where $E$ is Young's modulus, $\nu$ is the Poisson ratio (usually assumed to be close to $0.5$ for such gels \cite{Boudou2006}) and $\theta$ as defined above. 
$\delta_0$ is chosen so that the tip penetration depth into the sample is large enough to have a sufficient contact area and not too large to 
remain within the linear elasticity assumptions corresponding to the Hertz model. In order to carry out microrheology measurements, a small perturbation 
(frequency $f$ from $1\,\mathrm{Hz}$ to $0.5\,\mathrm{kHz}$, and $\omega=2\pi f$ is the angular frequency) was superposed onto the initial indentation. The 
perturbation being small, Eq. (\ref{indentation}) can be linearized about the equilibrium. 
By the correspondence principle of LVE, in the $\omega$--domain, one operates with complex quantities. Let $\delta^*$, $F^*$ be the complex indentation 
and force. Substracting the hydrodynamic drag $i\omega b(0)$ \cite{Alcaraz2003}, the complex shear modulus $G^*(\omega)$ is given by:
\begin{equation}
G^*(\omega)=\frac{1-\nu}{3\,\delta_0 \,\tan\theta} \{\frac{F^*(\omega)}{\delta^*(\omega)} - i\omega b(0)\}
\label{module}
\end{equation}

where $b(h)$ is a function which contains the geometry of the tip and depends on the height $h$ from tip to sample, and was measured 
as in \cite{Alcaraz2003} by extrapolation of the function at $h=0$. 

\begin{center}
\begin{figure}
\includegraphics[scale=0.8]{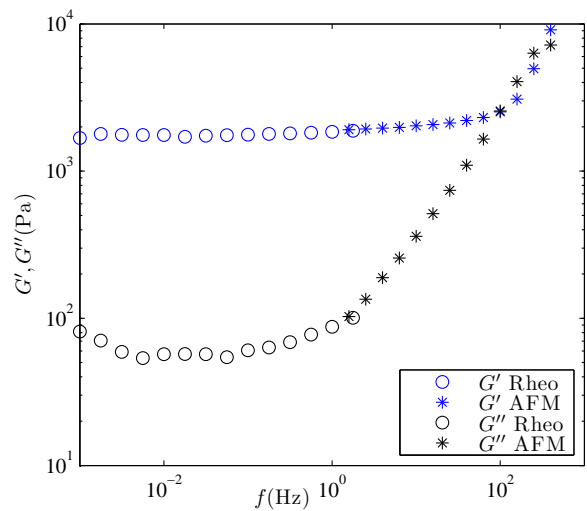}
\caption{\label{superposition} Superposition of rheometrical and AFM microrheological measurements. The acrylamide content is $10\%$ in this case. 
Typical error bars (not shown) are around $10\%$. $T=25\celsius$.}
\end{figure}
\end{center}

Rheometry measurements were carried out on a controlled stress rheometer (Malvern, Gemini 150) at low frequencies  
$[0.001\,\mathrm{Hz}-2\,\mathrm{Hz}]$ in the linear regime (deformation of $1\%$). Interestingly, an excellent agreement was found  
between these measurements and the AFM microrheology experiments for all gels, as seen in Fig.~\ref{superposition} where matching occurs 
around $2\,\mathrm{Hz}$ in the case of the $10\%$ concentration gel.

The variations of the dynamic moduli (Fig.~\ref{superposition}) show a constant elastic plateau modulus ($G_N^{0}$) at low frequencies ($G_N^{0}\simeq2300\,\mathrm{Pa}$ in 
Fig.~\ref{superposition}). The gel undergoes a glassy transition in the higher frequency regime as the AFM measurements do show. 
The slopes of the moduli $G'$ and $G''$ (around $1.0$) are similar above $100\,\mathrm{Hz}$.

To predict the observed behavior, a rheological model was used. The complex modulus $G^{*}(\omega)$ can be related to a relaxation 
function $H(\lambda)$ using the general formalism \cite{Baumgaertel1992} :
\begin{equation}
 G^{*}(\omega)=\int_0^\infty H(\lambda) \frac{i\omega \lambda}{1+i\omega \lambda}\frac{d\lambda}{\lambda}
 \label{integral}
\end{equation}
$H(\lambda)$ is the continuous relaxation spectrum, the expression of which is shown in this work to model the LVE response from flow 
to glassy state. In particular, the flow regime is described with the corresponding function $H_\mathrm{f}(\lambda)$ :
\begin{equation}
H_\mathrm{f}(\lambda) = 
    \left\{ 
       \begin{array}{ll}
          \displaystyle
          n_\mathrm{f} \,G_N^{0}\, \left(\frac{\lambda}{\lambda_\mathrm{max}}\right)^{n_\mathrm{f}} & \text{if } \lambda \le \lambda_\mathrm{max}
          \\
          \\
          0  & \text{if } \lambda >  \lambda_\mathrm{max}
       \end{array}
    \right.
    \label{formalism}
\end{equation}
%
%
This power law behavior will then describe the continuous relaxation time distribution required to model the plateau regime observed in 
Fig.~\ref{superposition} at low frequencies. This model is unsuitable to describe the high frequency state observed here. Another approach
based on the BSW description \cite{Baumgaertel1992} was found to be insufficient to predict the data accurately. Therefore a fractional 
derivative model \cite{Bagley-Torvick1983} is added to the previous one, to account for this behavior. 
The corresponding expression for the dynamic complex modulus $G_\mathrm{g}^*(\omega)$ is simply given by :
\begin{equation}
 G_\mathrm{g}^{*}(\omega)= G_1 \, \frac{(i \omega \lambda_1)^b}{1+(i \omega \lambda_1)^a}
\end{equation}
where $a$ and $b$ are the orders of fractional derivatives \cite{Bagley-Torvick1983}. Compatibility with thermodynamics requires 
$0 < a \leq b$ \cite{palade1996}. This type of model accounts for the slopes of the glassy transition, as observed in the current data. 

\begin{center}
\begin{figure}
\includegraphics[scale=0.8]{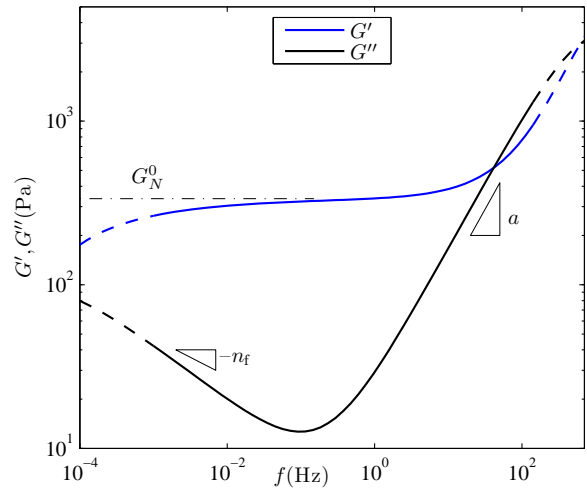}
\caption{\label{model} Significance of the model parameters $G_N^{0}$, $n_\mathrm{f}$, and $b=a$. $\lambda_\mathrm{max}$ is 
not shown but should appear at lower frequencies at the intersection of $G'$ and $G''$ occuring for $\omega \sim 1/\lambda_\mathrm{max}$. 
This does not occur in such gels since the flow region is not reached at low frequencies. Similarly $G_1$ corresponds to a higher plateau modulus in $G'$ 
but is well above our data, this being also the case for $\lambda_1$.}
\end{figure}
\end{center}

The coupling of the two linear models is insured by the simple relationship $G^*(\omega)=G^*_\mathrm{f}(\omega)+G^*_\mathrm{g}(\omega)$, 
to account for the whole frequency spectrum. The parameters of this global 
model are $G_N^{0}$, $\lambda_\mathrm{max}$, $n_\mathrm{f}$, $a$, $b$, $G_1$ and $\lambda_1$, . 
$G_N^{0}$ appears naturally to be the classical elastic plateau modulus (Fig.~\ref{model}). $\lambda_\mathrm{max}$ is the maximum relaxation time 
corresponding to the flow domain. In the case of gels, this specific time is out of reach since gels do not actually flow and 
exhibit a plateau even at very low frequencies \cite{verdier2009b}. $-n_\mathrm{f}$ is the slope of $G''$ 
at low frequencies, in a log--log plot and is found to lie between $-0.8$ and $0$.
$b$ represents the slopes of $G'$ and $G''$ in the glass transition regime (with $b=a$ in Fig.~\ref{model}). $G_1$ is the high frequency limit modulus of $G'$, 
and is far above our data, so complementary rheological experiments would be necessary to reveal such high values of the limiting modulus \cite{palade1996}. 
$\lambda_1$ is a time related to the microstructure and is not shown in Fig.~\ref{model}. Actually $1/\lambda_1$ is a typical crossover 
frequency between the glass transition and solid domain.
Finally $b-a$ would correspond to the limiting slope of $G'$ and $G''$ moduli at the highest frequencies, but is not shown in Fig.~\ref{model}. Here 
we used $a=b$, which was quite sufficient for describing our data, and led to the optimal fit.

Fitting of the data was carried out for the four gels characterized both in rheometry and AFM microrheology. 
The best-fitting values of the parameters were determined by minimizing a weighted sum of squared residuals. 
The weights were chosen from the data.
Minimization was achieved using the Levenberg--Marquardt method. The initial guesses followed the discussion on the role of each single 
parameter (see Fig.~\ref{model}).
The best-fitting values of the parameters are reported in Table~\ref{parameters} 
and the associated curves are presented below in Figures~\ref{Gels-all}. Very good agreement is shown.
\begin{figure}
\begin{tabular}{c}
\includegraphics[scale=0.7]{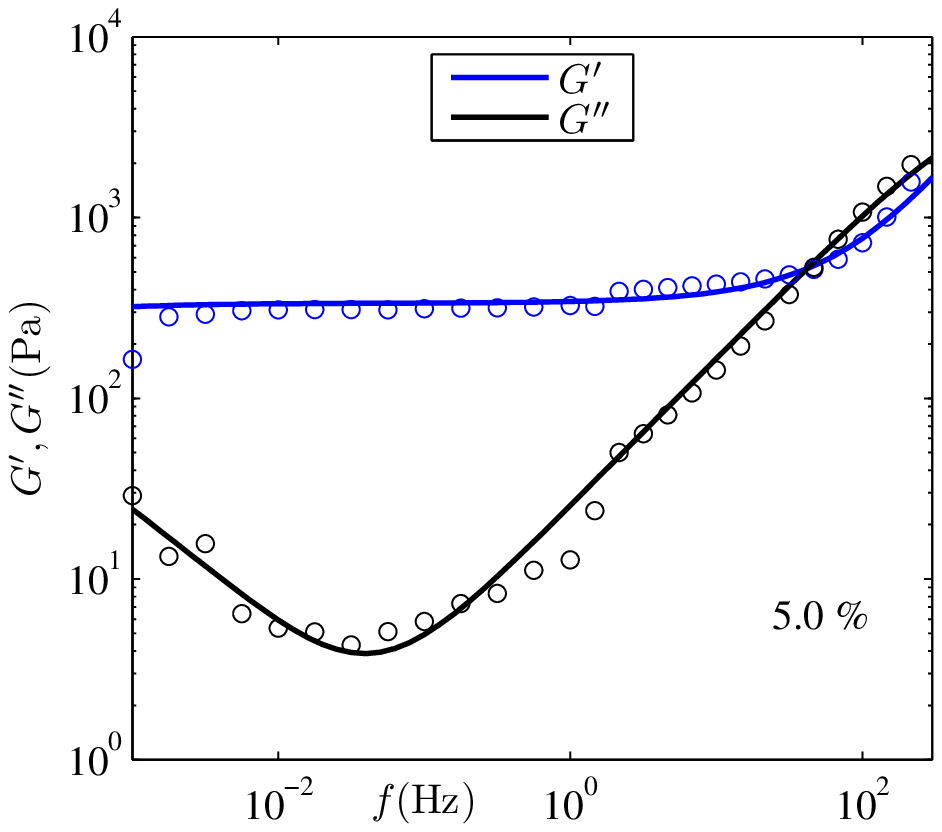}\\
\includegraphics[scale=0.7]{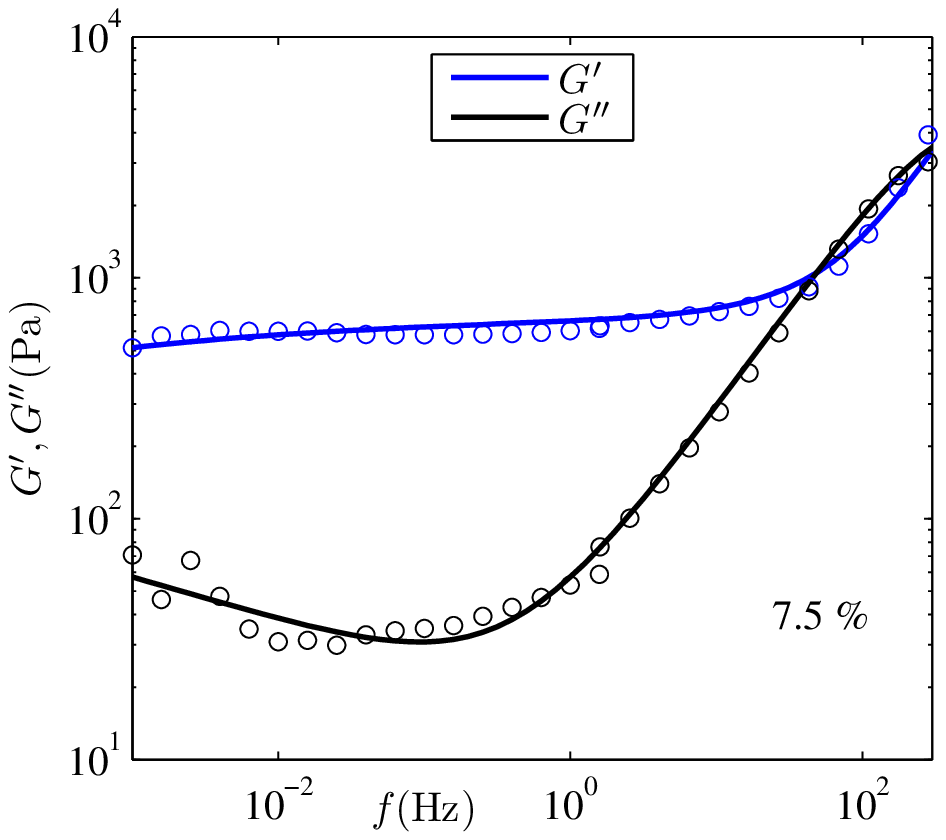}\\
\includegraphics[scale=0.7]{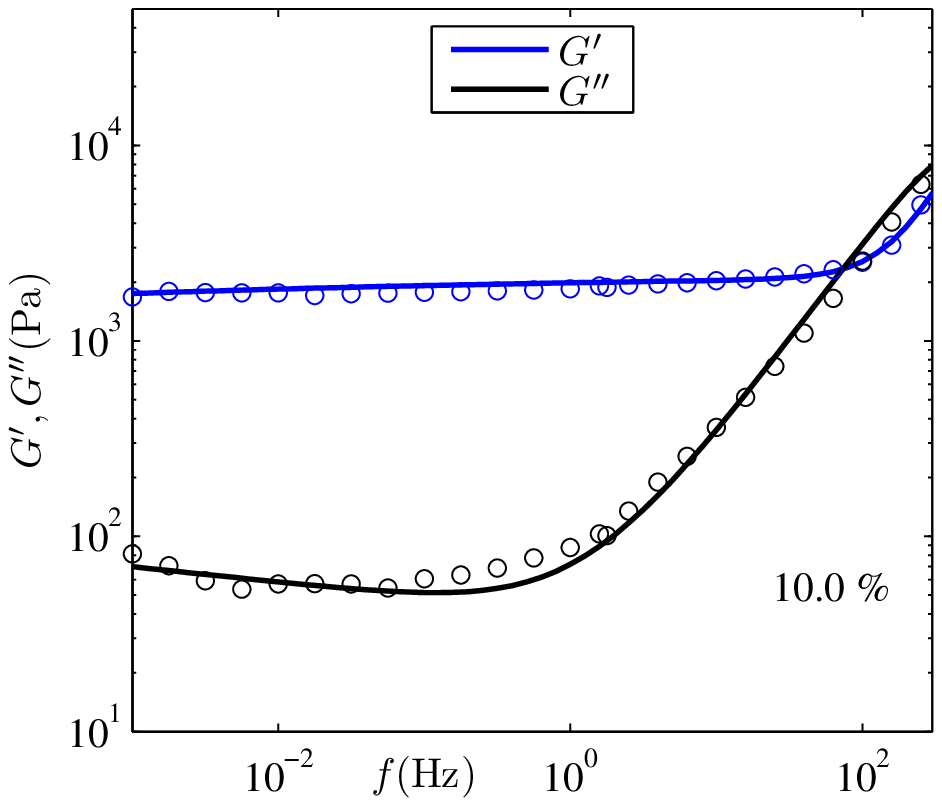}\\
\includegraphics[scale=0.7]{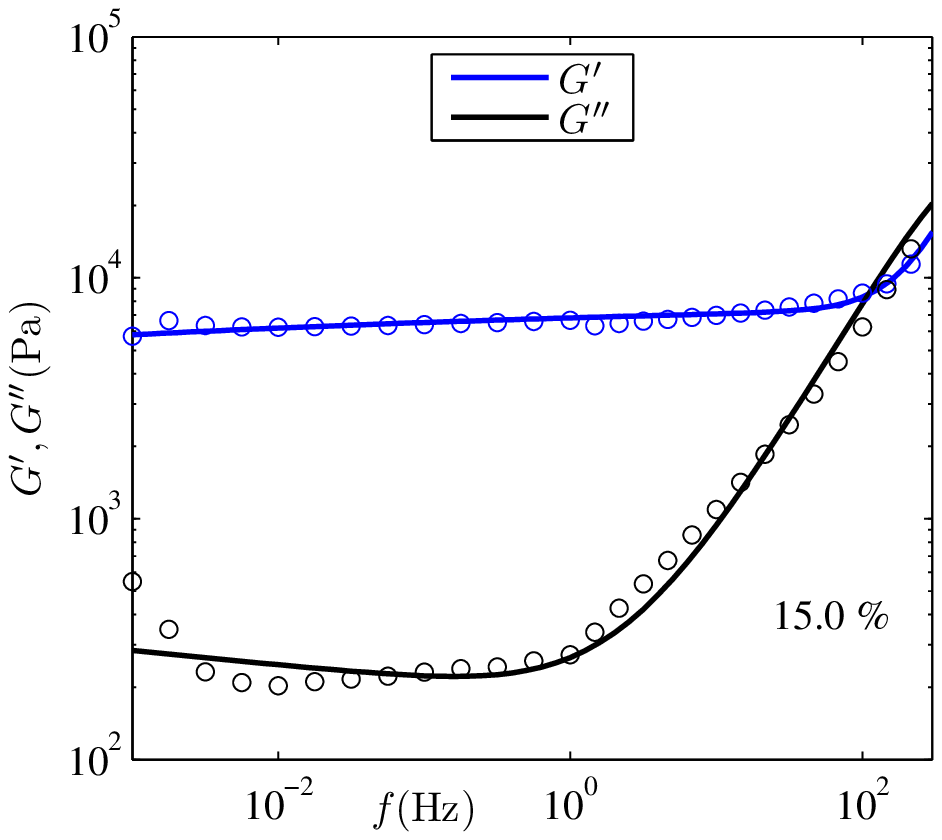}\\

\end{tabular}
\caption{\label{Gels-all} Gel rheology: 5\%, 7.5\%, 10\% and 15\% acrylamide concentrations. Open circles are experimental data whereas solid lines 
are the model best-fit curves.}
\end{figure}
\begin{table}
\setcellgapes{2pt}
\makegapedcells
\caption{\label{parameters} Best-fitting values of parameters used in the model.}
\begin{tabular}{lccccccr}
\hline
\hline
 Gel & $G_N^{0}(\mathrm{Pa})$ & $\lambda_\mathrm{max}(\mathrm{s})$ & $n_\mathrm{f}$ & $\lambda_1(\mathrm{s})$ & $G_1(\mathrm{Pa})$ &  $a=b$ \\
\hline
5\%      &  \numprint{336}   &  $1.5\times10^4$  &  0.73    &  $1.3\times10^{-4}$  & $1.0\times10^4$ &  0.82    \\
7.5\%    &  \numprint{710}   &  $2.0\times10^5$  &  0.18    &  $2.4\times10^{-4}$  & $1.0\times10^4$ &  0.85    \\
10\%     &  \numprint{2307}  &  $9.0\times10^9$  &  0.08    &  $2.4\times10^{-4}$  & $2.0\times10^4$ &  1.00    \\
15\%     &  \numprint{8801}  &  $1.0\times10^{10}$ &  0.06    &  $2.0\times10^{-4}$  & $5.9\times10^4$ &  1.00    \\
\hline
\hline
\end{tabular}
\end{table}

\begin{center}
\begin{figure}
\includegraphics[scale=0.7]{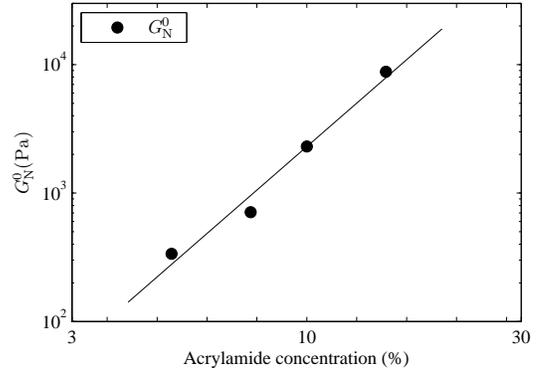}
\caption{\label{G0-G1} Evolution of gel moduli $G_N^{0}$ vs. concentration. The slope of the power law exponent corresponds to $3.0 \pm 0.3$.}
\end{figure}
\end{center}

Note that, as expected, the plateau modulus $G_N^{0}$ increases with $c$, the acrylamide concentration. These 
values are shown in Fig.~\ref{G0-G1} and the slope can be compared to other available data from the literature. For the lower frequency plateau, 
the relationship is of the kind $G_N^{0} \sim c^{3.0}$. Previous observations using combined light scattering and mechanical tests \cite{Hecht1978} 
were reported, showing an exponent $2.55$ using dynamic mechanical measurements (and $2.35$ using dynamic light scattering) as compared to the theory 
of de Gennes giving $2.25$ for good solvents \cite{degennes1976}. The value of the exponent for $G_N^{0}$ is also close to the exponent $2.55$, found for 
collagen gels \cite{iordan2010,Vader2009} but is larger than the typical exponent of $1.4$ obtained for entangled actin solutions \cite{hinner1998}.

The longest relaxation time $\lambda_\mathrm{max}$ does not seem to play a significant role, because it is related to a possible crossover 
of the $G'$ and $G''$ moduli at low frequencies which does not occur for such gels (in our frequency range) since they do not flow at 
low frequencies. However, $-n_\mathrm{f}$, the low--frequency slope of $G''$, is an important parameter here, and decreases as gel 
concentration increases. This further emphasizes the fact that high concentration gels exhibit moduli which have almost flat $G'$ and $G''$ 
moduli (see in particular Figures \ref{Gels-all} at 10\% and 15\% acrylamide concentration) and do not cross at low frequencies. Note that values 
of $G'$ and $G''$ at low frequencies ($0.001\,\mathrm{Hz}$) using 
the classical rheometry setup are difficult to obtain, due to the long experimental times required, therefore a larger uncertainty is 
unavoidable for $n_\mathrm{f}$. For the four gels, $\lambda_1$ was found to be almost constant within experimental error, as a function 
of concentration $c$, revealing no clear difference in the crosslinks relaxational processes of the acrylamide gels at very high frequencies. Therefore, 
crosslinks appear to link the acrylamide network in a regular mesh, and the short relaxation time 
$\lambda_1$ ($\sim 2.0\times10^{-4}s$) is gel independant.
This short time relaxation process corresponds to a single Maxwellian mode with values of $a$ and $b$ ranging between $0.82$ and $1.0$. 
In all cases, the optimal value of $b-a$ was found to be $0$, so we used $a=b$ in the optimisation method 
(see table \ref{parameters}). Remember that values of $a$ (or $b$) are directly related to the fractional derivatives present in the model.

Finally, the onset of the glass transition is an interesting parameter for such gels, and corresponds to a typical transition time $1/\lambda_T$. 
As can be seen in Figures \ref{Gels-all}, the corresponding frequency (around $100\,\mathrm{Hz}$) is increasing with polymer content, as curves are slightly 
shifted towards the right. 
The angular frequency $\omega_T=1/\lambda_T$ can be found analytically by investigation of the change of slope of $G'$. On the plateau $G' \sim G_N^{0}$ and 
at the transition $G' \sim G_1 \, cos(\pi a/2)(\omega \lambda_1)^a$ since $\omega \lambda_1 << 1$. Therefore the transition (angular) frequency 
$\omega_T$ is given by:

\begin{equation}
\omega_T = \frac{1}{\lambda_1}  \left(\frac{G_N^{0}}{G_1\,cos(\pi a/2)}\right)^{1/a}
\end{equation}

This can be checked easily from Table~\ref{parameters}. Indeed $\frac{G_N^{0}}{G_1}$ increases with polymer concentration since $G_N^{0}$ 
varies faster than $G_1$, while $cos(\pi a/2)$ decreases, thus the ratio $\frac{G_N^{0}}{G_1\,cos(\pi a/2)}$ increases and $1/a$ increases as well. 
Since $\lambda_1$ is almost constant, we deduce that $\omega_T$ increases with polymer concentration as shown experimentally. $\omega_T$ is linked to the 
ability of polymer crosslinks to move at such frequencies, and its increase shows that such motions are restricted due to the polymer excess 
at higher concentration (and constant crosslinker concentration). This typical frequency could be an important parameter to exhibit in 
future studies on gels.

Further extensions of the model may be considered for other physical (or chemical) gels, as well as the study of biological gels, involving cytoskeleton 
filaments such as actin, tubulin, and finally living cells \cite{Alcaraz2003}. Therefore, this model, coupled with the use of high frequency AFM measurements, 
allows to investigate different types of filamentous networks in order to determine their behavior in a large range of frequencies, and could be applied to 
investigate the microstructure of complex materials such as living cells. In particular the determination of the transition time ${\lambda_T}$ could be of 
importance.

\vspace{0.2cm}

We thank the ANR grant $n^{o}$ 12-BS09-020-01 (TRANSMIG), the French ministry of research for fellowship to Y. Abidine, the Nanoscience foundation 
for financial support of the AFM platform. The LIPhy laboratory is part of the LabEx Tec 21 (Investissements d$'$Avenir--grant agreement--$n^{o}$ANR-11-LABX-0030)".
Special thanks also go to E. Geissler for fruitful discussions on the rheology of polymeric gels.

\bibliographystyle{prsty}
\bibliography{gel}

\end{document}